\documentclass[10pt,conference]{IEEEtran}
\IEEEoverridecommandlockouts

\usepackage{cite}
\usepackage{amsmath,amssymb,amsfonts}
\usepackage{algorithmic}
\usepackage{graphicx}
\usepackage{textcomp}
\usepackage{xcolor}
\usepackage{graphicx}
\usepackage{multirow}
\usepackage{array}
\usepackage{setspace}
\usepackage{multirow, makecell}
\usepackage{rotating}
\usepackage{cleveref}
\usepackage{enumitem}

\def\BibTeX{{\rm B\kern-.05em{\sc i\kern-.025em b}\kern-.08em
    T\kern-.1667em\lower.7ex\hbox{E}\kern-.125emX}}

\begin{document}

\title{
Is Secure Coding Education in the Industry Needed? \\
An Investigation Through a Large Scale Survey
}

\author{\IEEEauthorblockN{Tiago Espinha Gasiba}
\IEEEauthorblockA{\textit{Siemens AG} \\
Munich, Germany \\
tiago.gasiba@siemens.com}
\and
\IEEEauthorblockN{Ulrike Lechner}
\IEEEauthorblockA{\textit{Universität der Bundeswehr}\\
\textit{München} \\
Munich, Germany \\
ulrike.lechner@unibw.de}
\and
\IEEEauthorblockN{Maria Pinto-Albuquerque}
\IEEEauthorblockA{\textit{Instituto Universitário de} \\
\textit{Lisboa (ISCTE-IUL), ISTAR-IUL} \\
Lisboa, Portugal \\
maria.albuquerque@iscte-iul.pt}
\and
\IEEEauthorblockN{Daniel Mendez}
\IEEEauthorblockA{\textit{Blekinge Institute of Technology} \\ \textit{and fortiss GmbH} \\
Karlskrona, Sweden\\
daniel.mendez@bth.se}
}

\maketitle

\begin{abstract}
The Department of Homeland Security in the United States estimates that 90\% of software vulnerabilities can be traced back to defects in design and software coding.
The financial impact of these vulnerabilities has been shown to exceed 380 million USD in industrial control systems alone.
Since software developers write software, they also introduce these vulnerabilities into the source code.
However, secure coding guidelines exist to prevent software developers from writing vulnerable code.
This study focuses on the human factor, the software developer, and secure coding, in particular secure coding guidelines.
We want to understand the software developers' awareness and compliance to secure coding guidelines and why, if at all, they aren't compliant or aware.
We base our results on a large-scale survey on secure coding guidelines, with more than 190 industrial software developers.
Our work's main contribution motivates the need to educate industrial software developers on secure coding guidelines, and it gives a list of fifteen actionable items to be used by practitioners in the industry. We also make our raw data openly available for further research.
\end{abstract}


\section{Introduction}

According to a Kaspersky~\cite{kaspersky_ics_2017} report, businesses spent an average of 380 million USD in 2017 to recover and deal with the consequences of Industrial Control Systems (ICS) incidents, and this value is still increasing.
Gartner's 2019 report predicts that the financial impact of attacks on Cyberphysical Systems will exceed 50 billion USD in 2023.
The United States Department of Homeland Security estimates that the vast majority of security incidents can be attributed to defects in software design and code~\cite{DHS_90_Percent}.

To deliver secure software-based products and services, we must consider security while producing software.
To become certified and able to conduct business in the critical infrastructure sector, companies must comply with several standards.
Among these standards, IT Security standards such as the ISO 27k~\cite{2013_27001} and IEC 62.443~\cite{2018_62443_4_1} mandate, among others, the establishment of a secure software development lifecycle (S-SDLC).
The S-SDLC includes the usage of secure coding guidelines (SCG) and the checking of code quality (ISO 25k~\cite{ISO250xx}) against these guidelines.

Secure coding, secure software development, and secure coding guidelines are no easy subjects.
Some of the vastly known and adopted SCG include Carnegie Mellon's Software Engineering Institute C, C++, and Java secure coding guidelines standars~\cite{CERT-SEI} (also known as SEI-CERT), the Motor Industry Software Reliability Association standard (MISRA)~\cite{MISRA_C:2012,MISRA_C_A1:2012}, and the Open Web Application Security Project Top 10 (OWASP standard)~\cite{owaspT10}.
However, SCG do not exist for all existing programming languages. 
In addition to SCG, to address the importance of secure code and the need to develop secure products, several companies united to form the SAFEcode~\cite{SAFECode}.
This alliance promotes secure coding and industrial secure coding best practices. 

Automatic tools such as Static Application Security Testing (SAST)~\cite{rodriguez2019software} can be used to automate and improve code quality.
These tools scan the code basis for existing vulnerabilities, which must be fixed by software developers.
However, previous research shows that their reliability is not good enough~\cite{oyetoyan2018myths}, in particular they exhibit a large amount of false positives and false negatives.
Also, these tools cannot automatically fix the code -- software developers must do this.

In this work, we focus on the human factor, i.e. the software developer. We justify this focus since it is the software developer who writes the code, who interprets the output of SAST tools, and who will ultimately be the person that introduces software weaknesses into the code basis. It will be the software developer as well that will have to correct the vulnerabilities in code.
This study is embedded in our investigation on the usage of serious games as a means to raise secure coding awareness of software developers in the industry \cite{gasiba_re19,Gasiba2020_Perliminary_Survey,Gasiba2020f,Gasiba2020_CyberICPS,Gasiba2020_CyberICPS_Journal,Gasiba2020c,Gasiba2019_Raising,Gasiba2020d,Gasiba2021_BSI}.
Our primary motivation to conduct the present work is to motivate awareness training by answering the question "is secure coding education in the industry needed?".
Our study focuses particularly on the education of secure coding guidelines.
We approach this question by looking at the perspective of software developers' compliance to secure coding guidelines.

Due to a lack of previous work exploring the relationship between secure coding guidelines and software developers' intention to comply with them, we have developed a survey to address this issue. Our previous publication details the overall research method underlying this survey and is available in~\cite{Gasiba2020_Perliminary_Survey}.
However, this previous publication focuses on the survey creation and only presents limited results from the survey pilot, i.e. it does not present any results of a large-scale deployment of the survey.
This work at hands closes this gap and presents an extensive analysis of a large-scale deployment.
We base our results on 194 answers from participants working in different industries, collected over a period of seven months. Our analysis of these results addresses the following research questions:

\begin{itemize}[leftmargin=+.435in]
    \item[ {\bf RQ1:} ] Which factors lead industrial software developers to comply with or ignore secure coding guidelines?
    \item[ {\bf RQ2:} ] To what degree are software developers aware of secure coding guidelines?
    \item[ {\bf RQ3:} ] To what extent is secure coding education in the industry needed?
\end{itemize}

Through the large-scale survey, our contribution to scientific knowledge comprises:
\begin{enumerate}
    \item openly available data from a large-scale survey, for other researchers to explore,
    \item the presentation and interpretation of results from the analysis of the survey, and
    \item a list of actionable items for practitioners and industrial cybersecurity educators.
\end{enumerate}

This paper is organized as follows.
Section~\ref{sec:related_work}, briefly discusses previous and related work.
Section~\ref{sec:survey} gives a very brief overview of the survey and its theoretical constructs.
In section~\ref{sec:results}, we present a comprehensive overview of the most important results from the analysis of the survey, derive actionable items, and discuss the threats to the validity.
This section, which constitutes the core of the paper, provides herein our main contribution.
Finally, section~\ref{sec:conclusion} concludes this paper with an overview of the study, and an outline of further work.

\section{Related Work}
\label{sec:related_work}

Based on a large-scale study by Patel et al.~\cite{gitlab_2019}, Bruce Schneier, a well-known security researcher, has stated that less than 50\% of software developers can spot security vulnerabilities in software~\cite{Schneier2020}.
Also, an estimation by the United States Department of Homeland Security, about 90\% of the {\it reported security incidents result from exploits against defects in the design or code of software}~\cite{DHS_90_Percent}.
Adding to these facts, software is becoming more complex and larger: a recent study by Sourcegraph~\cite{2020_Big_Code}, with more than 500 software developers, shows that more than 80\% of software developers are nowadays dealing with 20 times more code than ten years before.

An additional motivating factor for our work is Fisher et al.~\cite{fischer2017stack}, which shows that typical online platforms that software developers use to clarify development questions can be considered harmful.
The reason for not being a good source of information is that the answers present in these platforms are not curated in secure coding correctness. Their work indicates that severe problems can arise if software developers use these references and are not aware of secure coding practices.
Furthermore, Acar et al.~\cite{Acar2017} extensively analysed existing online resources that software developers can access to search about secure programming issues. They discovered that these platforms provide low-quality information in terms of cybersecurity. In particular, they found outdated information, wrong information, and no concrete examples or exercises.

While many studies focus on several different aspects of secure software development, very few empirical results exist on why software developers do not comply with secure coding practices. In particular, to the best of our knowledge, we have found no previous study addressing the aspects that lead industrial software developers to comply or not comply with secure coding guidelines in their daily work.
In a recent study Assal et al. \cite{Assal2019} analyzed how software developers are influenced and influence the secure coding processes.
They concluded that software developers are \textit{not the weakest link}, and are very motivated towards software security. However, they did not cover the reasons why this is so.
In 2011, Xie et al.~\cite{Xie2011} interviewed 15 senior professional software developers in the industry with an average of 12 years of experience. Their study shows a disconnect between software security concepts and the knowledge that the participants have in their jobs.
However, this study also does not focus on compliance to secure coding guidelines.

To address this issue, we have formally developed a survey \cite{Gasiba2020_Perliminary_Survey} to investigate software developers' compliance to secure coding guidelines. This survey is based on the adaptation of four distinct theories to the software developer context: IT Security Policy Compliance theory (PC), IT Security Neutralization theory (NT), Security-Related Stress theory (SRS), and IT Security Awareness (AW).
The work from Bulgurcu et al. \cite{bulgurcu2010information} and Moody et al. \cite{moody2018toward}) synthesizes the current research on IT security policy compliance.
Their work details the possible reasons that serve as factors for individuals to comply with IT security policies. Their constructs include, among others, the intention to comply and the knowledge of the policies.
Neutralization Theory is is discussed in \cite{siponen2010neutralization}, by Siponen et al., who address the possible reasons why subjects might find reasons to disregard IT security policies. Their constructs include, among others, the metaphor of the ledger, denial of injury, denial of responsibility, and appeal to higher loyalties.
D'Arcy et al. in \cite{d2014understanding}, discuss Security-Related Stress theory which uses coping theory to explore stress as a cause of deliberate IT security policy violations. Their constructs include, among others, the lack of understanding, higher workload, and constant changes.
Finally, Hänsch et al. provide a literature review on IT-security awareness \cite{2014_Benenson_Defining_Security_Awareness}.
Their conceptualization of IT Security Awareness comprises three distinct constructs: Perception, Protection, and Behavior. Perception relates to knowing existing threats, Protection relates to knowing existing mechanisms, and Behavior relates to actual behavior.
Finally, in this work, we use the recent results from WhiteSource~\cite{WhiteSource2019}, which present the top three vulnerabilities of the C, C++, Java, and Python programming languages. 

\section{Survey on Secure Coding Guidelines}
\label{sec:survey}

To investigate the possible reasons why vulnerabilities end up in final products, we have created a survey, described in \cite{Gasiba2020_Perliminary_Survey}, that focuses on industrial software developers.
This survey is based on the four established theories: policy compliance (PC, \cite{bulgurcu2010information,moody2018toward}), neutralization theory (NT, \cite{siponen2010neutralization}), security-related stress (SRS, \cite{d2014understanding}), and awareness (AW, \cite{2018_Graziotin_Happy_Developers}). Furthermore, the survey contains additional questions based on the industry experience by the first author. These questions are grouped by company background (CBG) and participant background knowledge (BGK).
\Cref{tab:survey:questions} shows the questions present in the survey, along with the different theories and constructs in which it is based.

The questionnaire comprises the following four sections: 1) demographic data, 2) secure coding awareness, 3) secure coding compliance, and 4) deterrents to compliance. 
The first part of the questionnaire includes general demographic questions on work experience, previous training on secure coding, the primary programming language used at work, used secure coding processes in the company, and software development method.
The second section of the questionnaire deals with awareness for secure coding. This part is individualized according to the primary programming language selected in the first section.
In this section, the participant is asked four questions related to high-impact vulnerabilities, according to~\cite{WhiteSource2019}.
These vulnerabilities are presented by the corresponding CWE~\cite{WWW_CWE_2019} description and number,
The four questions in this group correspond to Per1, Prot1, Be1 and BgK45, as shown in Table~\ref{tab:survey:questions}.
The answers to these (and only these) questions are based on a 3-point Likert scale: Yes, Uncertain and No.
The third section of the questionnaire presents questions to measure the intent to comply to secure coding guidelines. These are the questions marked with PC in Table~\ref{tab:survey:questions}.
Finally, the fourth section contains questions about the factors that influence compliance with secure coding guidelines.
These questions are based on neutralization theory and security-related stress, which are marked with NT and SRS in Table~\ref{tab:survey:questions}, respectively.

The answers to the questions are based on a 5-point Likert scale, which include the following: strongly disagree (SD), disagree (D), neutral (N), agree (A) and strongly agree (SA). The following mapping is used in our results: SD$\leftrightarrow$1, D$\leftrightarrow$2, N$\leftrightarrow$3, S$\leftrightarrow$4, SA$\leftrightarrow$5.
In the previous publication, we presented the rationale and details on how the survey was scientifically constructed. However, the preliminary results from the survey pilot available in this previous work only included a minimal subset of the survey questions, leading to very limited conclusions.
For more details on the survey' questions and the design of the survey, we refer the reader to \cite{Gasiba2020_Perliminary_Survey}.

\begin{spacing}{.9}
  \begin{table*}[http]
    \scriptsize
    \centering
    \caption{Survey Questions, Theories and Constructs}
    \label{tab:survey:questions}
    \resizebox{1.0\textwidth}{!}{
    \renewcommand{\arraystretch}{1.0}
    \begin{tabular}{|p{0.7cm}|p{0.5cm}|p{1.1cm}|p{11.7cm}|}
      \hline
      \textbf{Theory}    &
      \textbf{Ref.} &
      \textbf{Construct} &
      \textbf{~~~~~~~~~~~~~~~~~~~~~~~~~~~~~~~~~~~~~~~~~~~~~~~~~~~~~Survey Question}
      \\ \hline
      \hline
      
      \multirow{8}{*}{CBG} &
      \multirow{8}{*}{~~\textemdash} &
      \textit{CBg1} &
      In your company compliance to secure code guidelines is being checked in projects you work in
      \\ \cline{3-4}
      &
      &
      \textit{CBg2} &
      You know the secure software development lifecycle in your company
      \\ \cline{3-4}
      &
      &
      \textit{CBg3} &
      To which extent do you work with the $\rule{1cm}{0.15mm}$ secure coding standard? 
      \\ \cline{3-4}
      &
      &
      \textit{CBg4} &
      Could you explain why you use secure coding guidelines when writing code for the product you currently develop? 
      \\ \cline{3-4}
      &
      &
      \textit{CBg5} &
      Could you tell us why you do not use secure coding guidelines?
      \\ \cline{3-4}
      &
      &
      \textit{CBg6} &
      Why is compliance to secure coding guidelines not actively being checked in the projects you work in?
      \\ \cline{3-4}
      &
      &
      \textit{CBg7} &
      How is the compliance to secure coding guidelines checked in your current project?
      \\ \cline{3-4}
      &
      &
      \textit{CBg8} &
      In your company you use a well established secure software development life-cycle
      \\ \hline

      \multirow{5}{*}{BGK} &
      \multirow{5}{*}{~~\textemdash}&
      \textit{BgK1} &
      Compliance to secure coding guidelines is an important part of the development of company's products
      \\ \cline{3-4}
      &
      &
      \textit{BgK2} &
      Which of the following secure coding standards and best practices do you know? 
      \\ \cline{3-4}
      &
      &
      \textit{BgK3} &
      You are aware of negative consequences resulting from exploiting vulnerabilities in the products you work for
      \\ \cline{3-4}
      &
      &
      \textit{BgK4} &
      What other weaknesses do you pay attention to in developing software for the product you currently work for?
      \\ \cline{3-4}
      &
      &
      \textit{BgK5*} &
      You know about this weakness
      \\ \cline{1-4}

      \multirow{12}{*}{PC}&
      \multirow{6}{*}{~\cite{bulgurcu2010information}}&
      ISPA &
      You know that your company has a policy that mandates the usage of secure coding guidelines in software development
      \\ \cline{3-4}
      &
      &
      ITC &
      You intend to always comply with secure coding guidelines
      \\ \cline{3-4}
      &
      &
      GISA &
      You are aware of the existing security threats to the products of your company
      \\ \cline{3-4}
      &
      &
      SE-C1 &
      In your opinion, to write secure code, you have the necessary skills
      \\ \cline{3-4}
      &
      &
      SE-C2 &
      In your opinion, to write secure code, you have the necessary knowledge
      \\ \cline{3-4}
      &
      &
      SE-C3 &
      In your opinion, to write secure code, you have the necessary competency
      \\ \cline{2-4}
      &
      \multirow{2}{*}{~\cite{moody2018toward}}&
      FacCond5 &
      Support is available if you experience difficulties in complying with secure coding guidelines
      \\ \cline{3-4}
      &
      &
      RespCost4 &
      Secure coding guidelines make the task of writing software more difficult
      \\ \cline{2-4}
      &
      \multirow{4}{*}{~~\textemdash}&
      \textit{PC-Conf} &
      Complying to SCG makes you feel more confident about the security of the code that you write
      \\ \cline{3-4}
      &
      &
      \textit{PC-NT} &
      In your opinion, to write secure code, you have the necessary time
      \\ \cline{3-4}
      &
      &
      \textit{PC-NR} &
      In your opinion, to write secure code, you have the necessary resources
      \\ \cline{3-4}
      &
      &
      \textit{PC-NF} &
      In your opinion, to write secure code, you have the necessary freedom
      \\ \hline
      \multirow{11}{*}{NT} &
      \multirow{9}{*}{~\cite{siponen2010neutralization}} &
      N-DON3 &
      It is OK to disregard secure coding guidelines when this means that you deliver your work-packages faster
      \\ \cline{3-4}
      &
      &
      N-ATHL1 &
      It is OK to disregard secure coding guidelines when you would otherwise not get your job done
      \\ \cline{3-4}
      &
      &
      N-DOI1 &
      It is OK to disregard secure coding guidelines when this would result in no harm to the customer
      \\ \cline{3-4}
      &
      &
      N-DOI2 &
      It is OK to disregard secure coding guidelines if no damage is done to the company you work for
      \\ \cline{3-4}
      &
      &
      N-DOR3 &
      It is OK to disregard secure coding guidelines if you do not understand them
      \\ \cline{3-4}
      &
      &
      N-COC1 &
      It is not as wrong to ignore secure coding guidelines that are not reasonable
      \\ \cline{3-4}
      &
      &
      N-COC2 &
      It is not as wrong to ignore secure coding guidelines that require too much time to comply with
      \\ \cline{3-4}
      &
      &
      N-MOTL1 &
      You feel that your general adherence to secure coding guidelines compensates for occasionally ignoring them
      \\ \cline{2-4}
      &
      \multirow{3}{*}{~~\textemdash}  &
      \textit{NT-MArc} &
      It is OK to disregard secure coding practices when this would lead to major architectural changes
      \\ \cline{3-4}
      &
      &
      \textit{NT-CH} &
      It is OK to disregard secure coding guidelines when this means that it makes your company's customers happy
      \\ \cline{3-4}
      &
      &
      \textit{NT-SC} &
      It is OK to disregard secure coding guidelines if the software is not safety critical
      \\ \cline{1-4}
      \multirow{6}{*}{SRS}  &
      \multirow{6}{*}{~\cite{d2014understanding}}&
      CX2 &
      You find that new employees often know more about secure coding than you do
      \\ \cline{3-4}
      &
      &
      CX4 &
      You often find it difficult to understand your organization’s security coding guidelines
      \\ \cline{3-4}
      &
      &
      OL1 &
      Complying to secure coding guidelines forces you to do more work than you can handle
      \\ \cline{3-4}
      &
      &
      OL4 &
      You are forced to change your work habits to adapt to your organization’s secure coding guidelines
      \\ \cline{3-4}
      &
      &
      UC1 &
      There are constant changes in secure coding guidelines your organization
      \\ \cline{3-4}
      &
      &
      UC4 &
      There are constant changes in security-related technologies in your organization
      \\ \hline


      \multirow{3}{*}{AW}  &
      \multirow{3}{*}{~\cite{2014_Benenson_Defining_Security_Awareness}} &
      \textit{Per1*} &
      You can recognize code that contains this weakness
      \\ \cline{3-4}
      &
      &
      \textit{Be1*} &
      You know how to write code that does not contain this weakness
      \\ \cline{3-4}
      &
      &
      \textit{Prot1*} &
      You understand the possible consequences that can result from exploiting this weakness
      \\ \hline
    \end{tabular}}
    {\\\hspace{\textwidth}
                     {\scriptsize {\bf RQ.}: Research Question, {\bf CBG}: Company Background, {\bf BGK}: Participant Background Knowledge, {\bf PC}: Policy Compliance Theory, {\bf NT}: Neutralization Theory,~~~~~~~~~~~~~~~~~~~\\ {\bf SRS}: Security-Related-Stress Theory, {\bf AW}: Awareness, Note: constructs marked with * are specific for different programming languages}~~~~~~~~~~~~~~~~~~~~~~~~~~~~~~~~~~~~~~~~~~~~~~}
  \end{table*}
\end{spacing}

\subsection{A Large Scale Survey}
A large scale deployment of the survey was performed between March and September of 2020, resulting in a total running time of seven months.
The survey was announced through several different channels, in particular:
\begin{enumerate}
    \item {\bf Professional Contacts}: Linked-In, direct contacts by the authors at several different companies, Münchener Sicherheitsnetzwerk (Munich Security Network Forum)
    \item {\bf Social Media}: Twitter, Facebook, Reddit
    \item {\bf Other}: University contacts, survey exchange platform (SurveySwap), advertisement in university website
\end{enumerate}

The survey was constructed using the open-source survey platform LimeSurvey~\cite{LimeSurvey} Version 3.17.0+190402 and deployed in Amazon Web Services.
At the beginning of the survey, it was clearly stated: the purpose of the research, contact details, and the mandatory requirement that the participant must be a software developer from the industry.
Over the seven months, the survey was accessed 363 times resulting in 196 complete answers. Two answers were rejected due to irregularities found in the collected data. The full set of captured data is available under the following link~\cite{gasiba_zenodo_entire_survey}.
All the data was anonymously collected; however, a cookie was activated to prevent participants from submitting twice their answers.

\begin{figure}[http]
    \centering
    \caption{Survey demographics in terms of industry}
    \label{fig:survey:demographics:industry}
    \includegraphics[width=0.95\columnwidth]{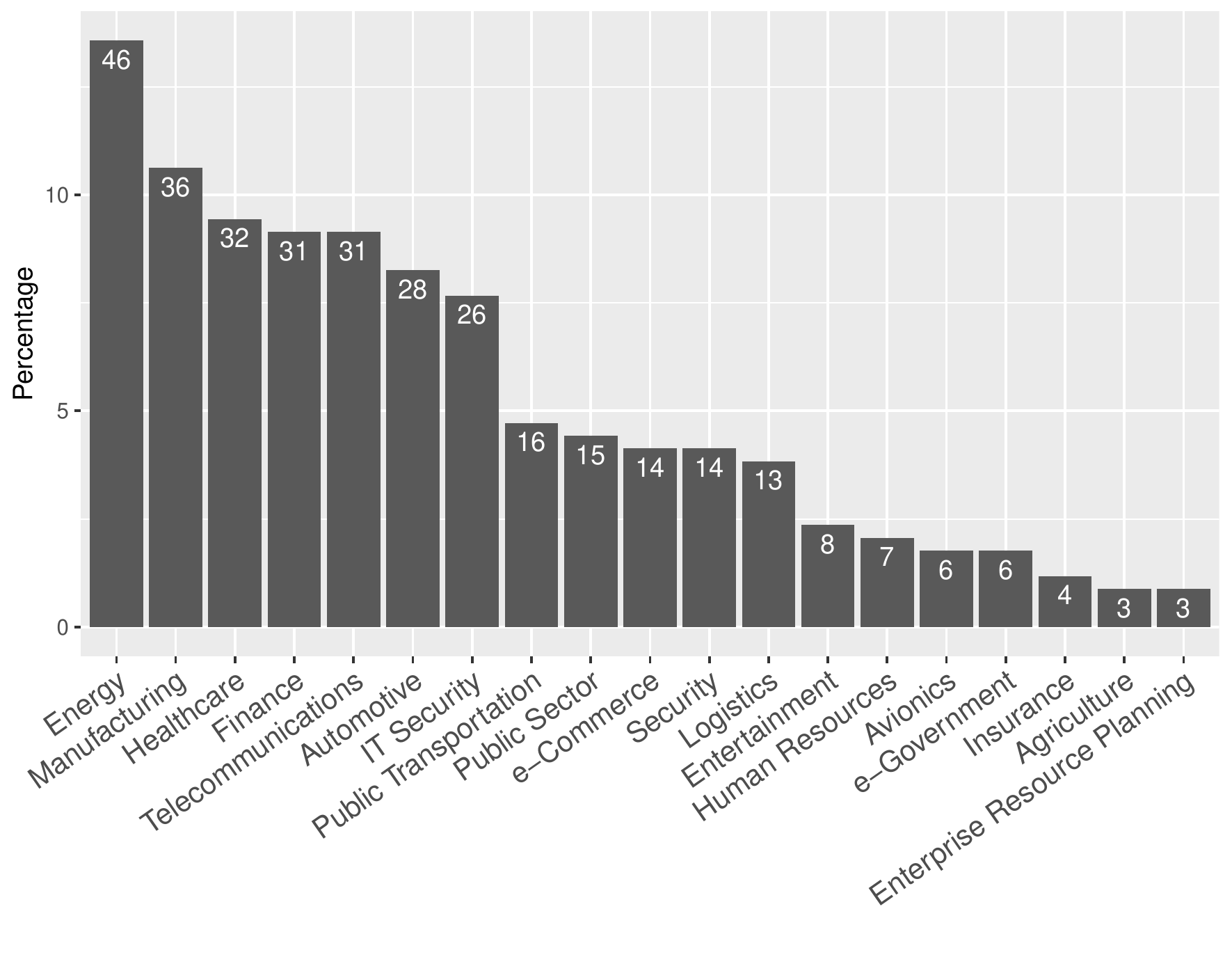}
\end{figure}

\Cref{fig:survey:demographics:industry} shows the different background industries captured by this data set. The demographics in terms of participants' programming languages are the following: C++ 50 (26\%), Java 38 (20\%), Python 37 (19\%), Other 36 (18\%), C 33 (17\%). The survey was anonymous, and geographical, education and gender aspects were not captured.

\vspace{-20pt}
In the next section, an extensive discussion of the results of the survey data analysis will be presented.

\section{Results}
\label{sec:results}

This section presents the survey results, categorized by the different theories in which it is based: CBG, BGK, PC, NT, SRS, and AW.
The section concludes with the main practical take-aways from the analysis and discusses the threats to the results' validity.

\subsection{Company Background}

\Cref{tab:CBg1_CBg2_CBg8_BgK1_BgK3} shows the results for the company background constructs CBg1, CBg2, and CBg8. From these results, we observe that, in general, {\it compliance to secure coding guidelines is not being checked in the industry}, and that software developers are not sure about the secure software development life-cycle (S-SDLC) used in their company. This observation is corroborated by the CBg8 results. 

\begin{table}[http]
    \centering
    \caption{Company Background (CBG) and Background Knowledge(BGK)}
    \label{tab:CBg1_CBg2_CBg8_BgK1_BgK3}
    \includegraphics[width=\columnwidth]{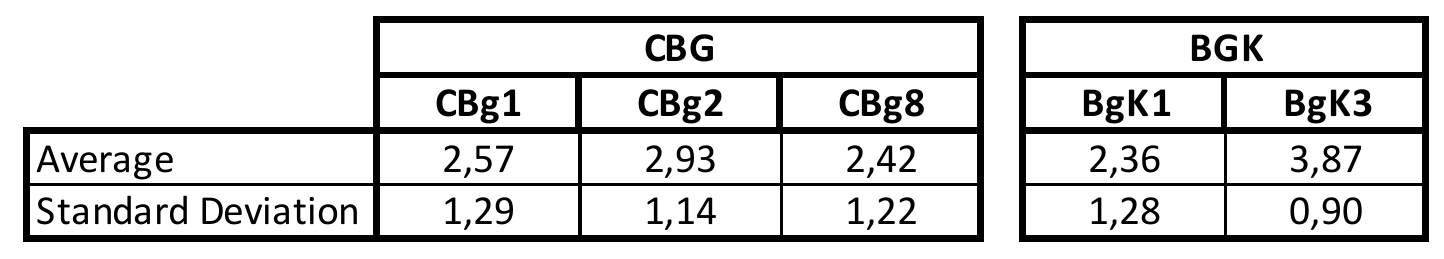}
\end{table}

\Cref{tab:CBg4_CBg5_CBg6} shows the results for CBg4, CBg5, and CBg6. Here we observe that about 50\% of the participants (out of the 23 that mentioned security being or not a requirement) claim that in their industry, implementation of security during product development is a requirement, while the other half state that this is not the case.
In terms of factors why SCG are not used (CBg5), we found the following important factors: 1) lack of awareness (focus on products and not on security, and limited knowledge), 2) relying on SAST tools, 3) because the participants had no previous experience with issues, and 4) limited or lack of management commitment, and resources devoted to security.   
The main factor not to check compliance to SCG (CBg6) is the fact that SCG is not used, and the industry focuses on products, not on security. Another important factor was the (perceived) lack of automatic tools to assist in the compliance checks and especially the lack of awareness.

\begin{table*}[http]
    \centering
    \caption{Results for Company Background: CBg4, CBg5, and CBg6}
    \label{tab:CBg4_CBg5_CBg6}
    \includegraphics[width=0.97\textwidth]{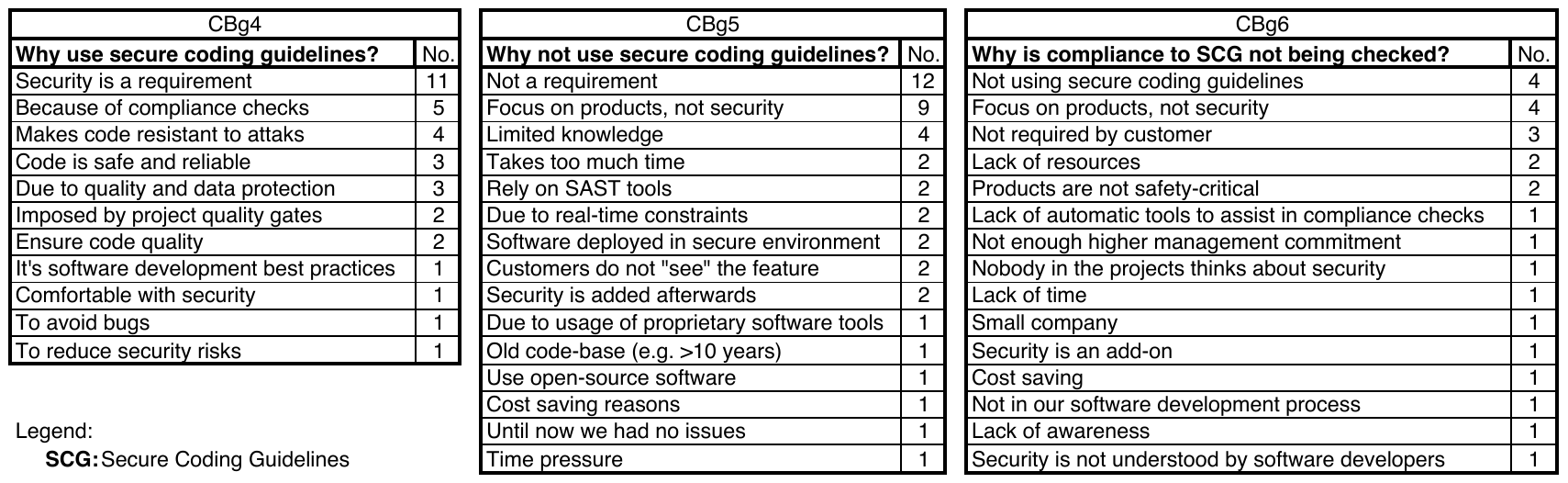}
\end{table*}

\begin{figure}[http]
    \centering
    \caption{CBg3: To which extent use secure coding standard}
    \label{fig:CBg3}
    \includegraphics[width=\columnwidth]{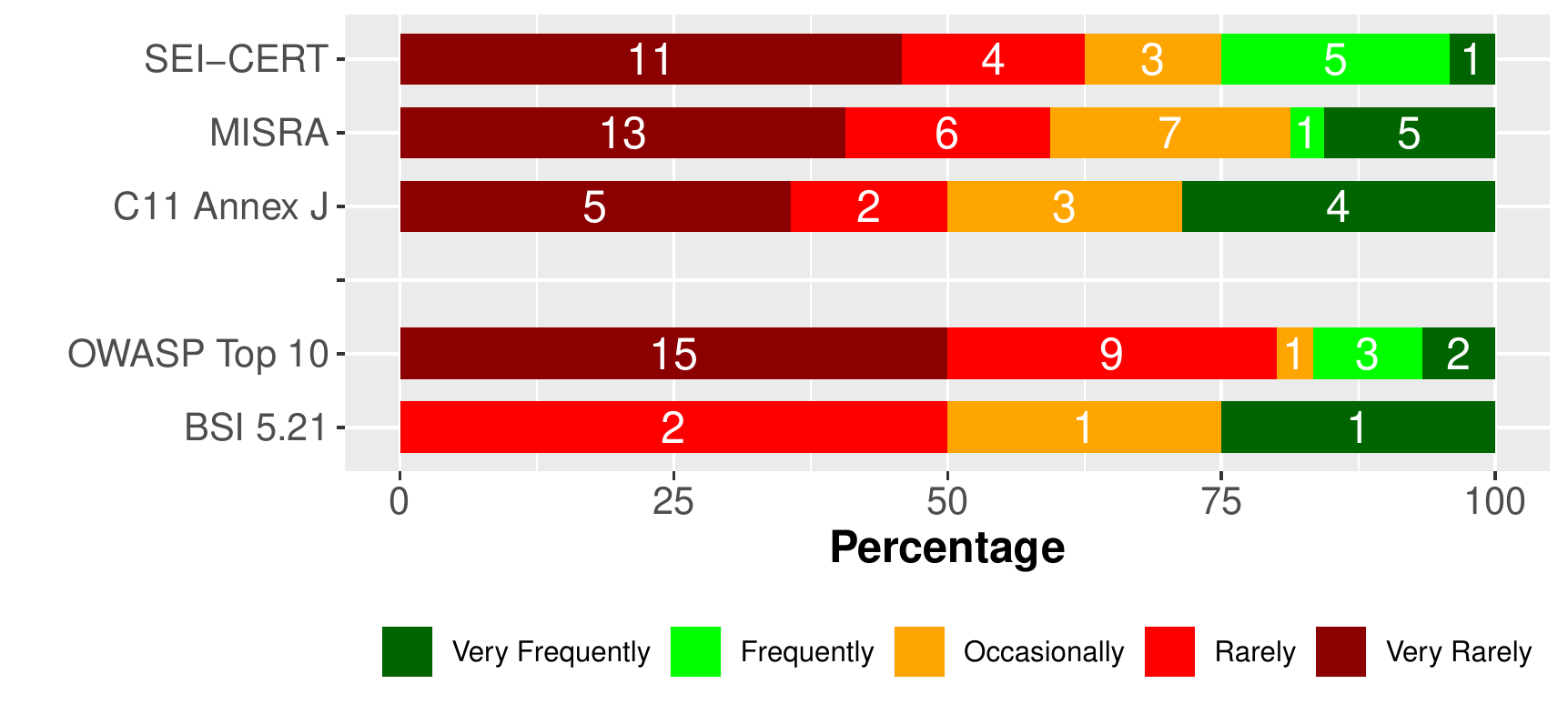}
\end{figure}

\Cref{fig:CBg3} shows the results of CBg3: to which extent are standard secure coding guidelines used in the industry. For all the secure coding standards that the survey has covered, all the results show that they are not really used in practice.

\begin{figure}[http]
    \centering
    \caption{CBg7: How are secure coding guidelines checked?}
    \label{fig:CBg7}
    \includegraphics[width=.9\columnwidth]{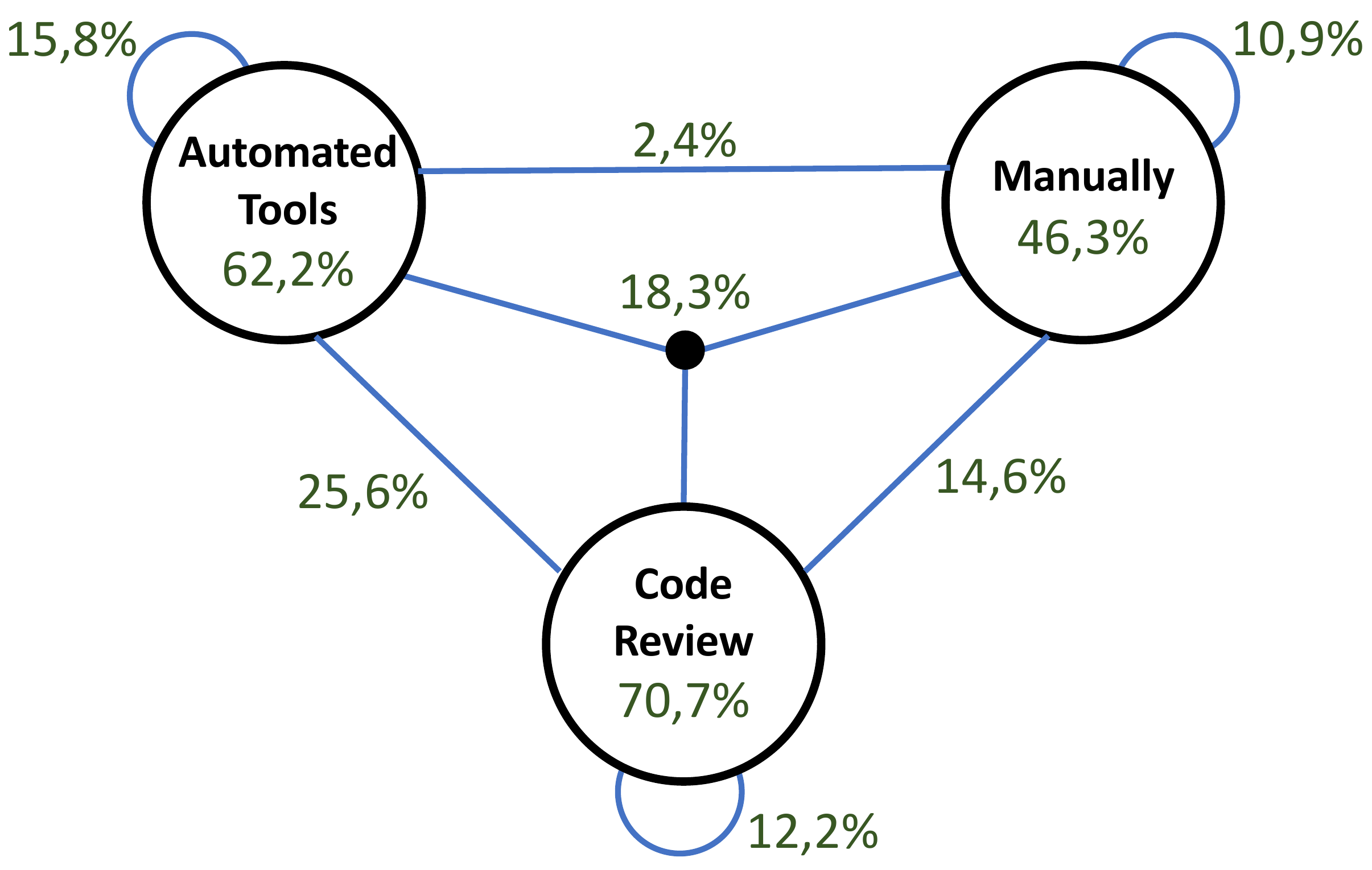}
\end{figure}

For the participants who answered that SCG are checked in their company, \cref{fig:CBg7} shows how they are being checked. This figure shows that 70.7\% check SCG during code review, 62,2\% using automated tools, and 46.3\% by a manual process. This figure also shows that 15.8\% of the checks are done using automated tools exclusively, 12.2\% using code review exclusively, and 10.9\% through a manual process. Employing two different methods (Automated Tools and Code Review) lead to 25.6\% of the results. About 18.3\% claim that the three methods are used simultaneously.

\subsection{Participant Background Knowledge}
\Cref{tab:CBg1_CBg2_CBg8_BgK1_BgK3} shows that compliance to secure coding guidelines (BgK1) is not considered an essential part of the development of products, with an average agreement of 2.36.
However, the survey participants have indicated to be aware (3.87 average agreement) of the negative consequences of exploiting software vulnerabilities (BgK3).
We attribute this observation to the large amount of advertisement, e.g., social media, on these negative consequences.

\begin{figure}[http]
    \centering
    \caption{BgK2: Knowledge of SCG standard}
    \label{fig:BgK2}
    \includegraphics[width=\columnwidth]{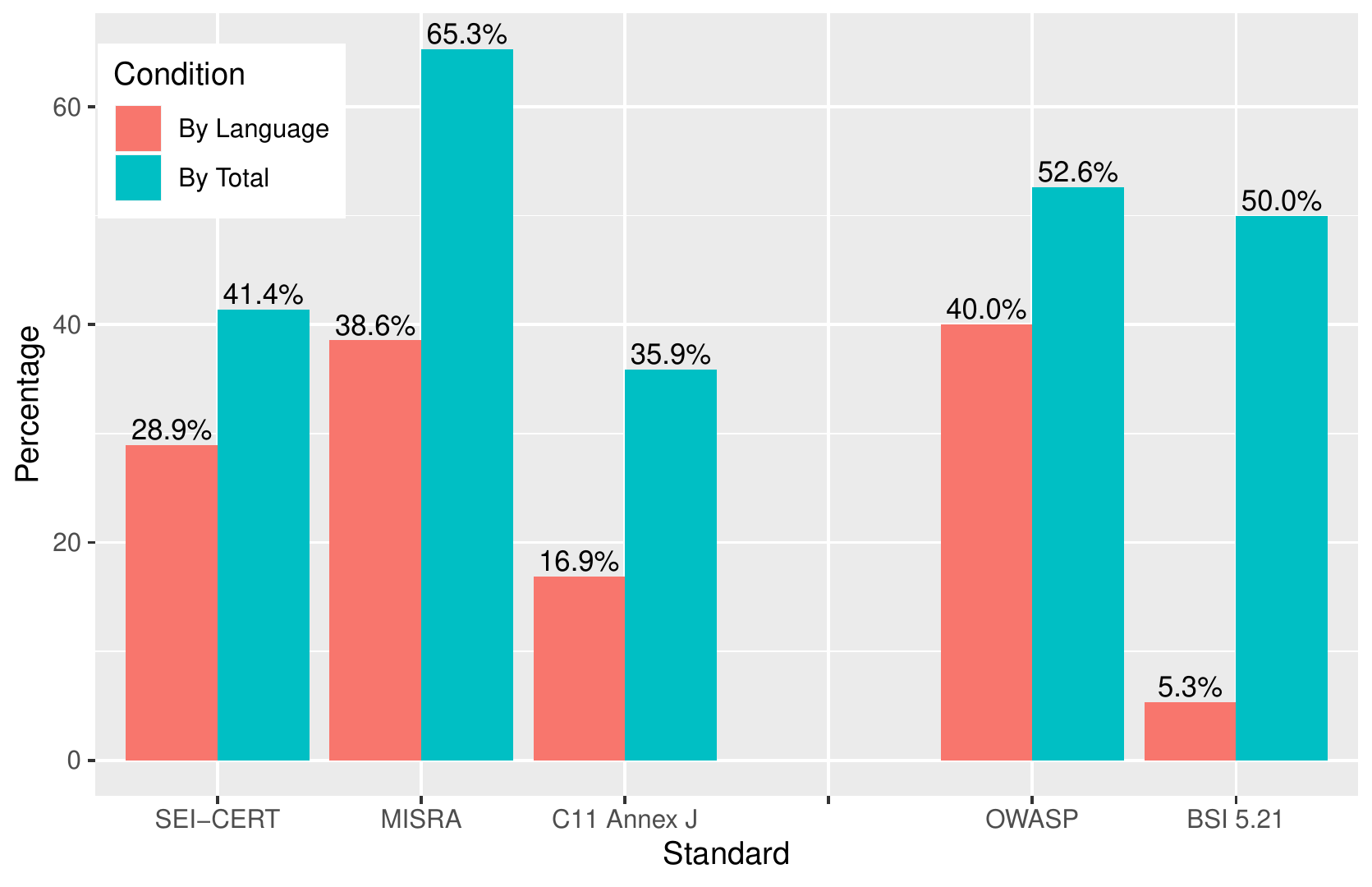}
\end{figure}

\Cref{fig:BgK2} shows the extent to which survey participants know the different secure coding standards.
The blue bars represent the "percentage of the total survey population that knows the given SCG standard". The red bars show the "percentage of the population that should know the standard, given their chosen programming language". For the SEI-CERT standard, this corresponds to the population who answered C or C++ as a programming language.
For the MISRA and C11 Annex J the results correspond to C programmers.
The OWASP and BSI standards capture Java, Python and programmers of Other languages.

\begin{table*}[http]
    \centering
    \caption{Results on PC, NT and SRS vs Industry, Programming Language and Work Experience}
    \label{tab:PC_NT_SRS}
    \includegraphics[width=.98\textwidth]{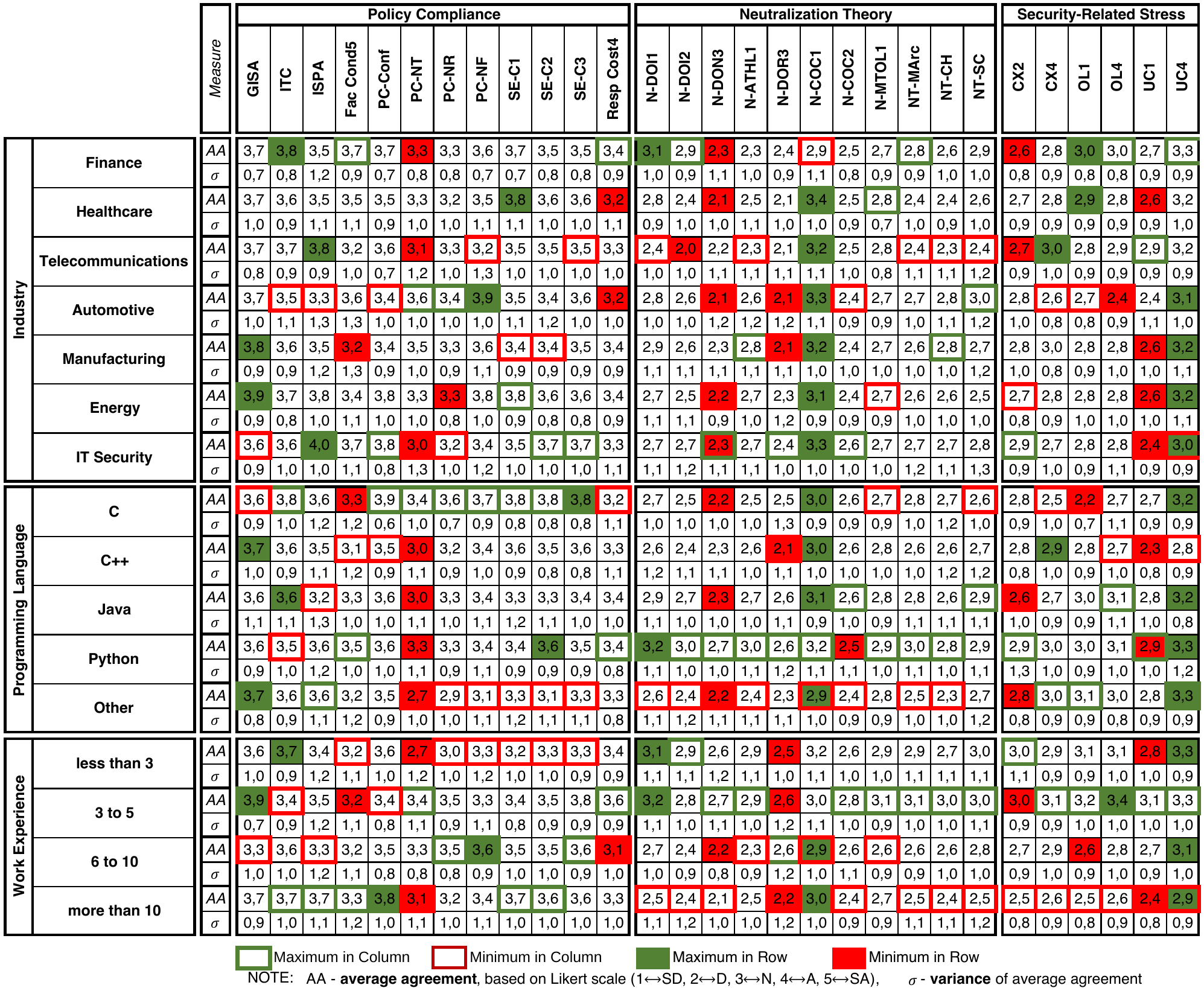}
\end{table*}

General knowledge about SCG is low (below 65.3\%). Comparing the blue to the red bars, these results show that the different standards are known to a larger percent of general population participants than those in the population that {\it should know the standard} -- this is an issue. It means that the population that should be more aware of these SCG standards is not aware of them. In particular, from the C and C++ software developers, only 28.9\% know the SEI-CERT standard, 38.6\% know the MISRA standard, and 16.9\% know the Annex J of the C11 standard. For developers using Python and Java, 40\% know the OWASP standard, and 5.3\% know the BSI 5.21 standard. This last result is not surprising, since the BSI standard is local to Germany only, and the survey was deployed on a global scale.

\subsection{Policy Compliance, Neutralization Theory, and Security-Related Stress}

\Cref{tab:PC_NT_SRS} shows the overall results for each theory (policy compliance, neutralization theory, and security-related-stress) for each theory construct, grouped by industry, programming language, and by work experience.
The minimum and maximum values are highlighted in this table, with the colors red and green, respectively. For a given group, the minimum and maximum in a column (i.e., per theory construct) is highlighted by a thicker border, while the minimum and maximum in a row is highlighted with a background color (red and green respectively).

In terms of policy compliance, we observe that the highest amount of agreement across all twelve constructs is obtained for the C programming language, while the highest amount of disagreement is obtained for other programming languages and participants with less than three years of industry experience.
We attribute the latter observation to the fact that newer employees need to accommodate to the job and might, therefore, not be yet fully integrated into the daily working life.
The construct that was rated with the lowest agreement across all the different groups is PC-NT (i.e. lack of time),
This result is to be expected due to the need to fulfill project deadlines in an industrial environment.

In terms of neutralization theory, the construct N-DON3 sees the largest amount of disagreement, i.e., software developers do not think that secure coding guidelines should be ignored to deliver work-packages faster.
However, there is a general agreement across all groups (industry, programming languages, and work experience) that ignoring unreasonable secure coding guidelines is acceptable.
This result is surprising since, according to the first author's experience, it is not the software developers' job to question the secure coding guidelines but comply with their policies when developing software.

Another surprising factor is that participants in the telecommunications industry find fewer reasons not to comply with secure coding guidelines.
According to the first author's experience, this might be because engineers working in this industry are used to developing software under tight constraints (e.g., real-time) and follow established coding guidelines to achieve this goal.
However, the IT security and finance department find more reasons not to comply with secure coding guidelines than other industries.
This fact is also surprising, especially for the IT security industry.
We think that, since the developers working in this industry face security topics daily, they might be more inclined to bend the established rules.
Another surprising factor is that, compared to the other programming languages, Python developers tend to find more reasons not to comply with secure coding guidelines.
We attribute this observation to the fact that Python is a prototyping language, where software developers might be more used to writing "quick and dirty" code, other than in the other cases.
Also, surprisingly, is the fact that software developers using programming languages other than C, C++, Java, or Python find fewer reasons not to comply with secure coding guidelines.
In terms of work experience, senior employees (more than ten years experience) tend to follow the established rules, while employees working for three to five years in the industry find more reasons to discard SCG.
Another result from this table is that, across all the groups, software developers also tend to ignore SCG that they do not understand.

Finally, in terms of security-related-stress, there is a general agreement on the construct UC4, i.e., the participants to the survey have observed constant changes in security-related technologies.
This observation might be related to the large and growing amount of different software development frameworks and changing (agile) software development methodology.
However, there is also a general disagreement on UC1, i.e., that secure coding guidelines are not continually changing.
We find this last observation positive since constantly changing secure coding guidelines can lead to unnecessary stress at work.

\subsection{Awareness}

For each of the programming languages, the participants were asked to answer Yes, Unsure, and No on how they agree with each of the awareness constructs (Per1, Prot1, Be1), and also on BgK5.
Additionally, each of these questions was asked in relation to a top-3 CWEs (Common Weakness Enumeration) that affects the programming language, according to the study by WhiteSource~\cite{WhiteSource2019}.
\Cref{fig:Awareness} shows the survey results for these constructs, for each programming language and each CWE.
We note that each of the CWE is related to one or more secure coding guideline~\cite{Gasiba2020c}.
In this table, an "unsure" answer was considered a negative aspect, therefore combined with "no" results.

The survey participants report high levels of awareness for BgK5 (knowing the vulnerability), and for the construct Prot1 (understanding the consequences of exploiting vulnerabilities). However, for Per1 (ability to recognize vulnerable code) and Be1 (knowing how to write secure code), the levels of awareness are low (less than 51\%). The first result is in line with the study by Patel et al.~\cite{gitlab_2019}, however, the second result is new in this study.
For the construct Per1, we also observe that the programming languages "Other", Python and C are especially at risk since the awareness level is low for their ranked vulnerabilities.

Considering all the constructs together, we observe an overestimation (60\% vs 40\%) of the participants' awareness level, since real-world data shows that the number of incidents is increasing.
We attribute this to {\it optimism bias}~\cite{thaler2009nudge}, which is a well-known effect in risk perception that occurs when someone overestimates or underestimates risk while remaining ignorant about their poor assessment~\cite{lechner2019security}.
Our results indicate an overestimation bias, which is corroborated with the industry's experience from the first author.

\begin{table*}[http]
    \centering
    \caption{Awareness Results vs Programming Language}
    \label{fig:Awareness}
    \includegraphics[width=\textwidth]{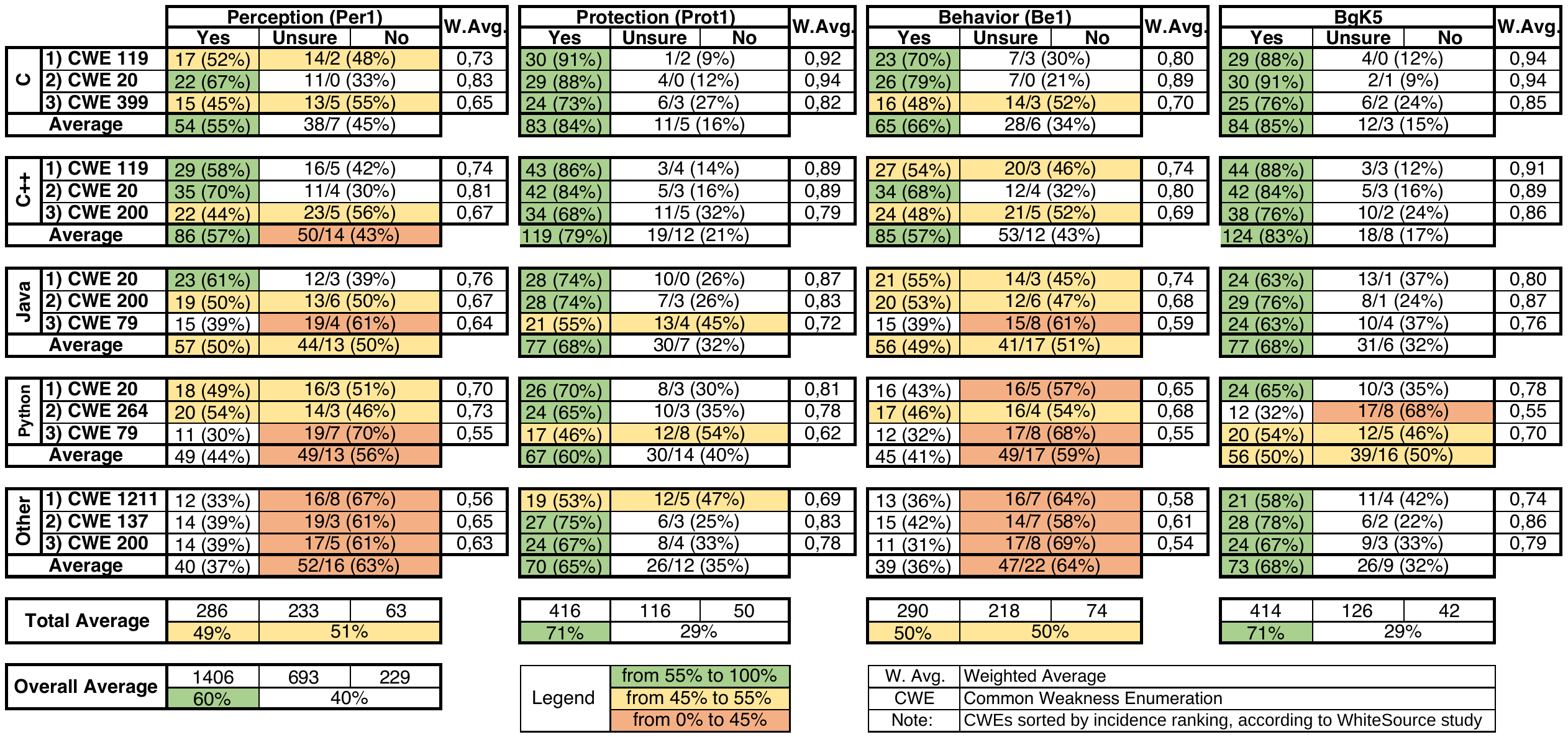}
\end{table*}

Since the participants were only asked to rank the top-3 CWE, in BgK4 we asked the participants to optionally name additional weaknesses that they pay attention to while developing software.
The participants' answers were coded to separate the correctly identified weaknesses from the vulnerabilities and general issues not related to secure coding.
\Cref{tab:BgK4} shows the result of the codification of the answers given by the participants.

\begin{table*}[http]
    \centering
    \caption{BgK4: Additional knowledge on coding weaknesses}
    \label{tab:BgK4}
    \includegraphics[width=\textwidth]{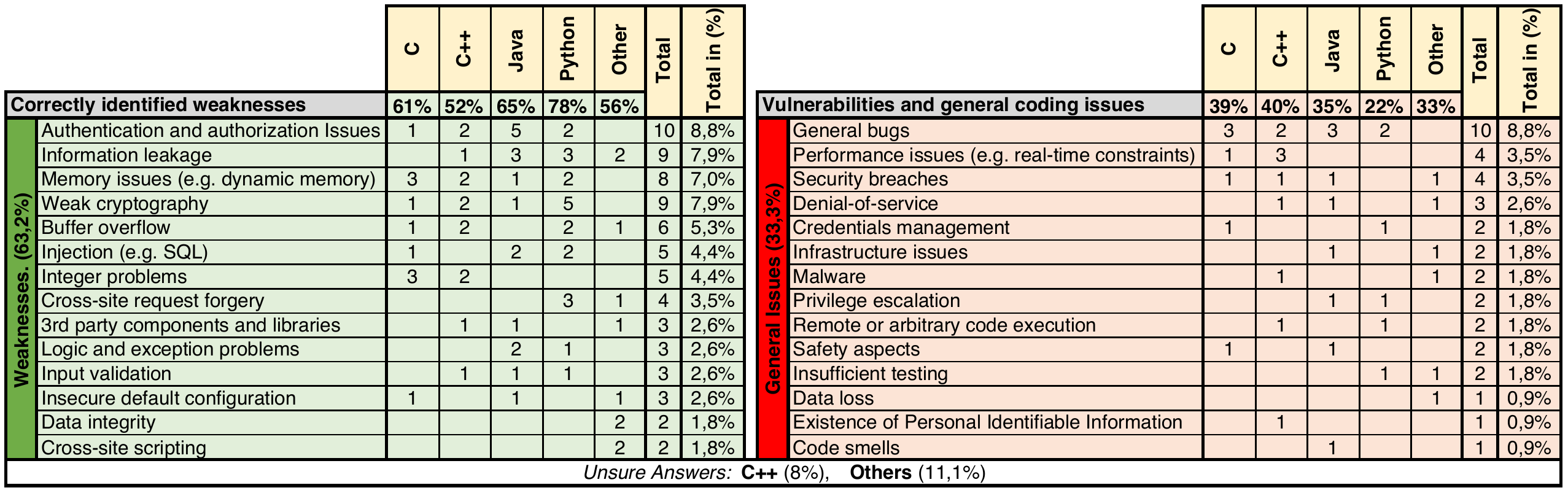}
\end{table*}

From all the additional survey answers, 63.2\% are software weaknesses, 33.3\% are general coding issues, and 3.5\% of the answers are from unsure participants.
Python has the highest amount of correctly identified weaknesses and, surprisingly, C++ the least amount.
The reason for the last observation might be due to the complexity of the C++ language.
In terms of correctly identified issues, Authentication and Authorization, Information leakage, Memory Issues, Weak Cryptography, and Buffer Overflow are in the top-5.
Surprisingly, the survey participants have considered general bugs, performance issues, and security breaches as secure coding weaknesses.
Denial-of-service is generally not considered a coding issue but a deployment issue (solved with e.g., load balancing). However, it was also considered a software vulnerability, ranking in the top-5 of the general issues category.
Surprisingly, also considered as secure coding issues have been: infrastructure issues, safety aspects, and code smells, personal identifiable information, and lack of testing.
In particular, code smells, which are {\it symptoms of poor design and implementation choices that may hinder code comprehensibility and maintainability}~\cite{palomba2018diffuseness}, have been shown to be generally dissociated from security vulnerabilities~\cite{elkhail2019relating}.
The consideration of these factors as secure coding weaknesses leads us to notice the lack of awareness of secure coding guidelines.

\subsection{Actionable Items for Industrial Practitioners}
In the following, we present the main actionable items (AI) we infer from the current work, i.e. from the analysis of the survey results but also from our experiences surrounding this topic in industrial application.
These actionable items should be taken into consideration by practitioners.
We split them into two main categories: general issues and secure coding guidelines. The AIs under the general issues category are not directly related to secure coding guidelines and include:
\begin{enumerate}
    \item {\bf Need to involve management}: without management understanding and approval, it is not possible to establish secure coding practices in a company
    \item {\bf Need to improve knowledge on company's internal S-SDLC and secure coding policies}: the survey has shown that software developers are not always aware of the company's internal policies and about the S-SDLC; therefore, specialized internal campaigns should be started to raise awareness of these issues
    \item {\bf Raise awareness of the difference between secure coding and other aspects, e.g. safety and performance}: the survey has shown that software developers tend to confuse these topics. When training software developers, the difference between these aspects should be made clear, as also possible opposing recommendations
    \item {\bf Consider security in the requirements phase}: it is no surprise that security should be considered early in the software development phases; rarely will customers "ask for security"; however, they will expect secure products. Therefore, company policies should be adapted to cover security from early stages, and software developers should be aware of the necessary steps to take (e.g., threat and risk analysis, secure architecture, security requirements).
\end{enumerate}

Furthermore, software developers must be made aware of the difference between security weakness and security vulnerability. In particular, security weaknesses are coding errors that might lead to a vulnerability, while (according to ENISA~\cite{ENISA_Glossary}, definition G52), a security vulnerability is the existence of a security weakness that can lead to a security breach. In terms of secure coding guidelines, we conclude the following key AIs for practitioners:

\begin{enumerate}
    \item {\bf Include SCG as an integral part of S-SDLC}: SCG should be lived as a process and should be second nature to software developers; daily practice and usage has the potential to have long-lasting and beneficial effects.
    \item {\bf Build a secure coding community}: promote secure coding practices inside the company. Some possible ways to implement this it to join larger communities which also promote secure coding practices, e.g., SAFEcode~\cite{2018_SAFEcode}, have monthly or weekly presentations or discussions on a secure coding topic, promote and use secure coding gamification (e.g., best secure coder of the month).
    \item {\bf Define a responsible person}: have a point-of-contact for secure coding issues; the job of this person includes making sure that software developers are trained, and motivated. 
    \item {\bf Implement awareness training on SCG}: a substantial amount of software developers needs training on SCG; as such, awareness training events should be promoted and held regularly.
    \item {\bf Implement hands-on awareness training}: motivated by the observed optimism bias, and also experience from the industry, we think that an effective way to raise awareness is to challenge the knowledge of software developers on secure coding topics; while desired training methods was not captured through the survey, our experience has shown that after being challenged on these topics, many developers tend to re-evaluate their knowledge and seek more information. A good way to achieve this is by the usage of Capture-the-Flag events, which are specially designed to raise awareness of secure coding for software developers in the industry; these exercises should mainly focus on the defensive perspective but also cover offensive aspects; furthermore, these exercises should provide a good motivation on why certain SCG exist -- this way, software developers can develop a better knowledge of SCG and understanding on why they should comply to them.
    \item {\bf Implement SCG quality gates}: secure code is also high-quality code; practitioners should consider adding the requirement of checking secure coding guidelines to typical project quality gates; some ways this can be achieved include using specialized tools that are configured to check secure coding guidelines, keeping track of code reviews including review of secure coding guidelines, and status monitoring of software security testing results.
    \item {\bf Do not use SAST as a replacement for SCG}: some participants to the survey have mentioned the usage of SAST tools as a replacement for a formal training or consideration of secure coding guidelines; we consider this an important hidden danger -- previous studies~\cite{aloraini2019empirical,li2020vulnerabilities} have reported on the poor quality of SAST tools; therefore, we conclude that the human factor cannot be taken out of the loop and SAST tools should only be used in a supportive role.
    \item {\bf Monitor the quality of SAST tools}: since the quality of the output of these tools might be poor, strategies to address their quality needs to be considered; in particular, the secure coding champion needs to implement a process to verify the quality of the tools being deployed and to replace them when outdated; additionally, if possible, it should be considered to use several tools in parallel, in order to compare the different results.
    \item {\bf Training on SCG should focus on concepts, not specific cases or instances}: the results of the survey indicate constant changes in security technologies; this might be related to the vast amount of existing frameworks and booming IT security field; these changes can cause unnecessary stress to software developers; as such, when dealing with secure coding guidelines, software developers should be trained on concepts and not so much on particular instances of SCG, in particular, software developers should understand the underlying reason for the SCG and not just assume the rule without further consideration.
    \item {\bf Keep up-to-date with the latest technology}: software developers should be informed about the latest security technologies, when necessary, especially when starting new projects; this should be done taking into consideration that too much information can cause stress, while too little information can mean that important news are missed; for this reason, we propose that the secure coding champion should constantly monitor new technologies and decide on their importance and introduction on running projects.
    \item {\bf Adapt SCG, only when necessary}: similar to keeping up-to-date with the latest technology, secure coding guidelines should be updated regularly; however, without interfering with ongoing projects; updates of SCG should be timed together with other SCG awareness campaigns, e.g., awareness training.
\end{enumerate}

\subsection{Threats to Validity}
In this work, we present the analysis of a large scale anonymous survey on the usage of secure coding guidelines in the industry, including a total number of 194 participants distributed across the globe.
Since the survey took place in an online format, and the collected data is anonymized, it is impossible to control the respondents' true background. However, we have counter-acted this possible bias in two different ways: by making sure that the channels where the survey was announced included a rich set of industrial software developers and that this requirement was clearly stated at the start of the survey (in particular with the following sentence in the beginning of the survey: "this survey is for software developers working in the industry").
Geographical background was not captured, which might impact our conclusions. Additionally, the different industry sectors are not equally represented, which might introduce bias to our conclusions.
Our conclusions result from an interpretation of the data in light of the first author's own experience in the industry. However, these results have been discussed with three additional security experts, whereby the conclusions hereby presented have been confirmed by all.
We observe that the survey results display optimism bias. To counterbalance this effect, we focus our results, not on absolute values, but on a relative comparison between different values.
Finally, the results on knowledge of the BSI standard might have a strong bias, since this is a local standard to Germany, and geographic results are not available.

\subsection{Impact of this work}
The results presented in this work provide an impact both in the academic community but also in the industry.
It is not always easy to obtain a large volume of survey data from participants from the industry.
In this work, we have collected the survey answers from over 190 industrial software developers.
This means that we can assume that our results have a strong supportive basis.
Also, the survey that was administered to the participants underwent an extensive design cycle, to make sure that it is based on well established scientific theories.
As a result of the analysis of the survey data, fifteen key take-away messages were derived, which serve both as guidance for future scientific work but also as valuable information for industry practitioners.
Our work not only addresses the awareness of software developers on the topic of secure coding guidelines, but it also raises awareness in the scientific community on this difficult topic.
Additionally, we provide the raw survey data, as a means to contribute to further research on the topic.

\section{Conclusions}
\label{sec:conclusion}
Cybersecurity is becoming ever more important nowadays.
Ignoring cybersecurity can lead to severe financial penalties or even loss of certification, together with loss of business or even loss of life, in critical infrastructures.
However, the last years have seen an increase in cybersecurity incidents.
According to an estimate by the United States Department of Homeland Security, the root cause of about 90\% of security incidents can be traced back to software design and coding weaknesses.
Secure coding guidelines exist to make software secure -- compliance to them increases the security and quality of code.
Additionally, Static Application Security Test tools also exist to reduce software vulnerabilities; however, previous studies have shown that these tools exhibit many false positives and false negatives.
These facts lead us to the following questions: 1) how aware are software developers of secure coding guidelines, 2) is secure coding education in the industry needed, and which factors lead software developers to comply or ignore secure coding guidelines.

In this work, we address these questions through a large-scale survey on software developers in the industry.
The survey design, which is a complicated endeavor by itself, is addressed in a separate publication, while this focuses on the analysis of the results and practical aspects and advice for practitioners and cybersecurity educators in the industry.
Our measurement of policy compliance is based on three established theories: policy compliance theory by Bulgurcu et al. and Moody et al.; neutralization theory by Siponen et al.; and security-related stress theory by D'Arcy et al.
Our measurement of awareness is based on the three dimensions, as defined by Hänsch et al.: perception, protection, and behavior.

Our results indicate that a large amount of software developers are not aware of secure coding guidelines. Previous studies argue that increasing awareness leads to increased compliance.
Therefore, we conclude that a method to address this lack of awareness is through education on secure coding.
Based on our results and experience, we also infer a set of fifteen actionable items for practitioners and industrial cybersecurity educators.
A further contribution herein is the raw survey results, which we make openly available for further research.
In future work, we would like to practice the derived actionable items and investigate novel methodologies for secure coding education of industrial software developers. We want to use these actionable items to improve our ongoing action-design research in the industry.

\section*{Supporting Data}
The raw data collected in the survey is openly available in Zenodo~\cite{gasiba_zenodo_entire_survey}.
The raw survey data is provided in Comma Separated Values (CSV) format.
Researchers are encouraged to make use of this data for further work.

\section*{Acknowledgements}
The authors would like to thank the participants of the survey for their constribution.
Also, the authors would like to thank Kristian Beckers  and Thomas Diefenbach for their helpful, insightful, and constructive comments and discussions.

This work is financed by portuguese national funds through FCT - Fundação para a Ciência e Tecnologia, I.P., under the project FCT UIDB/04466/2020. Furthermore, the third author thanks the Instituto Universitário de Lisboa and ISTAR-IUL, for their support.

\bibliographystyle{IEEEtran}
\bibliography{bibliography}

\begin{thebibliography}{10}
\providecommand{\url}[1]{#1}
\csname url@samestyle\endcsname
\providecommand{\newblock}{\relax}
\providecommand{\bibinfo}[2]{#2}
\providecommand{\BIBentrySTDinterwordspacing}{\spaceskip=0pt\relax}
\providecommand{\BIBentryALTinterwordstretchfactor}{4}
\providecommand{\BIBentryALTinterwordspacing}{\spaceskip=\fontdimen2\font plus
\BIBentryALTinterwordstretchfactor\fontdimen3\font minus
  \fontdimen4\font\relax}
\providecommand{\BIBforeignlanguage}[2]{{%
\expandafter\ifx\csname l@#1\endcsname\relax
\typeout{** WARNING: IEEEtran.bst: No hyphenation pattern has been}%
\typeout{** loaded for the language `#1'. Using the pattern for}%
\typeout{** the default language instead.}%
\else
\language=\csname l@#1\endcsname
\fi
#2}}
\providecommand{\BIBdecl}{\relax}
\BIBdecl

\bibitem{kaspersky_ics_2017}
\BIBentryALTinterwordspacing
{Kaspersky}, ``{The State of Industrial Cybersecurity -- 2017},'' 2017.
  [Online]. Available: \url{https://tinyurl.com/y7dfppak}
\BIBentrySTDinterwordspacing

\bibitem{DHS_90_Percent}
\BIBentryALTinterwordspacing
{Department of Homeland Security, US-CERT}, ``{Software Assurance},'' Sep.
  2020. [Online]. Available: \url{\url{https://tinyurl.com/y6pr9v42}}
\BIBentrySTDinterwordspacing

\bibitem{2013_27001}
{ISO 27001}, ``{Information technology -- Security techniques -- Information
  security management systems -- Requirements},'' {International Standard
  Organization}, Geneva, CH, Standard, Oct. 2013.

\bibitem{2018_62443_4_1}
{IEC 62443-4-1}, ``Security for industrial automation and control systems -
  part 4-1: Secure product development lifecycle requirements,'' {International
  Electrotechnical Commission}, Standard, Jan 2018.

\bibitem{ISO250xx}
\BIBentryALTinterwordspacing
{ISO}, ``{ISO 250xx Series},'' International Organization for Standardization,
  Geneva, CH, Standard, 2005. [Online]. Available:
  \url{http://iso25000.com/index.php/en/iso-25000-standards}
\BIBentrySTDinterwordspacing

\bibitem{CERT-SEI}
\BIBentryALTinterwordspacing
{Carnegie Mellon University}, ``{SEI-CERT Coding Standards}.'' [Online].
  Available: \url{\url{https://wiki.sei.cmu.edu/confluence/display/seccode}}
\BIBentrySTDinterwordspacing

\bibitem{MISRA_C:2012}
{--}, ``{Guidelines for the use of the C language in critical systems},'' Motor
  Industry Software Reliability Association, Nuneaton, Warwickshire, UK,
  Standard, Mar 2012.

\bibitem{MISRA_C_A1:2012}
------, ``{Additional security guidelines for MISRA C:2012},'' Motor Industry
  Software Reliability Association, Nuneaton, Warwickshire, UK, Standard, Mar
  2016.

\bibitem{owaspT10}
\BIBentryALTinterwordspacing
``{OWASP Top 10},'' Jul. 2017. [Online]. Available:
  \url{\url{https://tinyurl.com/yyb8wcv9}}
\BIBentrySTDinterwordspacing

\bibitem{SAFECode}
\BIBentryALTinterwordspacing
{SAFECode Charter Members}, ``{SAFECode - Software Assurance Forum for
  Excellence in Code},'' accessed Mar. 2020. [Online]. Available:
  \url{https://safecode.org}
\BIBentrySTDinterwordspacing

\bibitem{rodriguez2019software}
M.~Rodriguez, M.~Piattini, and C.~Ebert, ``{Software verification and
  validation technologies and tools},'' \emph{{IEEE Software}}, vol.~36, no.~2,
  pp. 13--24, 2019.

\bibitem{oyetoyan2018myths}
T.~D. Oyetoyan, B.~Milosheska, M.~Grini, and D.~S. Cruzes, ``{Myths and facts
  about static application security testing tools: an action research at
  Telenor digital},'' in \emph{{International Conference on Agile Software
  Development}}.\hskip 1em plus 0.5em minus 0.4em\relax Springer, Cham, 2018,
  pp. 86--103.

\bibitem{gasiba_re19}
\BIBentryALTinterwordspacing
T.~Gasiba, K.~Beckers, S.~Suppan, and F.~Rezabek, ``{On the Requirements for
  Serious Games geared towards Software Developers in the Industry},'' in
  \emph{Conference on Requirements Engineering Conference}, D.~E. Damian,
  A.~Perini, and S.~Lee, Eds.\hskip 1em plus 0.5em minus 0.4em\relax {Jeju,
  South Korea}: IEEE, Sep. 2019, pp. 286--296. [Online]. Available:
  \url{https://ieeexplore.ieee.org/xpl/conhome/8910334/proceeding}
\BIBentrySTDinterwordspacing

\bibitem{Gasiba2020_Perliminary_Survey}
T.~Gasiba, U.~Lechner, M.~Pinto-Albuquerque, and D.~M. Fernandez, ``{Awareness
  of Secure Coding Guidelines in the Industry - A first data analysis},'' in
  \emph{TrustCom 2020: International Conference on Trust, Security and Privacy
  in Computing and Communications}.\hskip 1em plus 0.5em minus 0.4em\relax
  {Guangzhou, China}: {IEEE}, Dec. 2020.

\bibitem{Gasiba2020f}
T.~Gasiba, U.~Lechner, M.~Pinto-Albuquerque, and A.~Zouitni, ``{Design of
  Secure Coding Challenges for Cybersecurity Education in the Industry},''
  \emph{{13th International Conference on the Quality of Information and
  Communications Technology, QUATIC}}, pp. 223--237, 09 2020.

\bibitem{Gasiba2020_CyberICPS}
T.~Gasiba, U.~Lechner, M.~Pinto-Albuquerque, and A.~Porwal, ``{Cybersecurity
  Awareness Platform with Virtual Coach and Automated Challenge Assessment},''
  in \emph{6th Workshop On The Security Of Industrial Control Systems \& Of
  Cyber-Physical Systems (CyberICPS)}.\hskip 1em plus 0.5em minus 0.4em\relax
  Online: Springer, Cham, 12 2020, pp. 67--83.

\bibitem{Gasiba2020_CyberICPS_Journal}
T.~Gasiba, U.~Lechner, and M.~Pinto-Albuquerque, ``{Sifu - A CyberSecurity
  Awareness Platform with Challenge Assessment and Intelligent Coach},'' in
  \emph{Cybersecurity Journal, Special Issue on Cyber-Physical System
  Security}.\hskip 1em plus 0.5em minus 0.4em\relax SpringerOpen, 12 2020, pp.
  1--23.

\bibitem{Gasiba2020c}
T.~Gasiba, U.~Lechner, J.~Cuellar, and A.~Zouitni, ``{Ranking Secure Coding
  Guidelines for Software Developer Awareness Training in the Industry},'' in
  \emph{First International Computer Programming Education Conference (ICPEC
  2020)}, ser. OpenAccess Series in Informatics (OASIcs), R.~Queirós,
  F.~Portela, M.~Pinto, and A.~Simões, Eds., vol.~81.\hskip 1em plus 0.5em
  minus 0.4em\relax Dagstuhl, Germany: Schloss Dagstuhl--Leibniz-Zentrum für
  Informatik, 2020, pp. 11:1--11:11.

\bibitem{Gasiba2019_Raising}
T.~Gasiba and U.~Lechner, ``{Raising Secure Coding Awareness for Software
  Developers in the Industry},'' in \emph{2019 IEEE 27th International
  Requirements Engineering Conference Workshops (REW)}.\hskip 1em plus 0.5em
  minus 0.4em\relax {Jeju, South Korea}: IEEE, Sep. 2019, pp. 141--143.

\bibitem{Gasiba2020d}
T.~Gasiba, U.~Lechner, F.~Rezabek, and M.~Pinto-Albuquerque, ``{Cybersecurity
  Games for Secure Programming Education in the Industry: Gameplay Analysis},''
  in \emph{First International Computer Programming Education Conference (ICPEC
  2020)}, ser. OpenAccess Series in Informatics (OASIcs), R.~Queirós,
  F.~Portela, M.~Pinto, and A.~Simões, Eds., vol.~81.\hskip 1em plus 0.5em
  minus 0.4em\relax Dagstuhl, Germany: Schloss Dagstuhl--Leibniz-Zentrum für
  Informatik, 2020, pp. 10:1--10:11.

\bibitem{Gasiba2021_BSI}
T.~Gasiba, U.~Lechner, and M.~Pinto-Albuquerque, ``{CyberSecurity Challenges:
  Serious Games for Awareness Training in Industrial Environments},'' 2 2021,
  {in Bundesamt für Sicherheit in der Informationstechnik (Hg.): Deutschland.
  Digital. Sicher. 30 Jahre BSI – Tagungsband zum 17. Deutschen
  IT-Sicherheitskongress}.

\bibitem{gitlab_2019}
\BIBentryALTinterwordspacing
S.~Patel, ``{2019 Global Developer Report: DevSecOps finds security roadblocks
  divide teams},'' Jul. 2020. [Online]. Available:
  \url{https://tinyurl.com/y6oypsh3}
\BIBentrySTDinterwordspacing

\bibitem{Schneier2020}
B.~Schneier, ``{Software Developers and Security},'' {Online}, Jul. 2020,
  https://www.schneier.com/blog/archives/2019/07/software\_develo.html.

\bibitem{2020_Big_Code}
\BIBentryALTinterwordspacing
Sourcegraph, ``{The Emergence of Big Code -- A 2020 Survey of Software
  Professionals},'' Oct 2020. [Online]. Available:
  \url{https://tinyurl.com/y5yfprn8}
\BIBentrySTDinterwordspacing

\bibitem{fischer2017stack}
F.~Fischer, K.~B{\"o}ttinger, H.~Xiao, C.~Stransky, Y.~Acar, M.~Backes, and
  S.~Fahl, ``{Stack overflow considered harmful? the impact of copy\&paste on
  android application security},'' in \emph{{2017 IEEE Symposium on Security
  and Privacy (SP)}}, {IEEE}.\hskip 1em plus 0.5em minus 0.4em\relax {an Jose,
  CA}: {IEEE}, 2017, pp. 121--136.

\bibitem{Acar2017}
Y.~Acar, C.~Stransky, D.~Wermke, C.~Weir, M.~L. Mazurek, and S.~Fahl,
  ``Developers need support, too: A survey of security advice for software
  developers,'' in \emph{2017 IEEE Cybersecurity Development (SecDev)}.\hskip
  1em plus 0.5em minus 0.4em\relax {Cambridge, MA, USA}: IEEE, Sep. 2017, pp.
  22--26.

\bibitem{Assal2019}
H.~Assal and S.~Chiasson, ``{'Think secure from the beginning' A Survey with
  Software Developers},'' in \emph{Proceedings of the 2019 CHI Conference on
  Human Factors in Computing Systems}, ser. CHI ’19.\hskip 1em plus 0.5em
  minus 0.4em\relax New York, NY, USA: Association for Computing Machinery,
  2019, pp. 1--13.

\bibitem{Xie2011}
J.~Xie, H.~R. Lipford, and B.~Chu, ``{Why do Programmers Make Security
  Errors?}'' \emph{2011 IEEE Symposium on Visual Languages and Human-Centric
  Computing (VL/HCC)}, pp. 161--164, Sep. 2011.

\bibitem{bulgurcu2010information}
B.~Bulgurcu, H.~Cavusoglu, and I.~Benbasat, ``{Information Security Policy
  Compliance: An Empirical Study of Rationality-Based Beliefs and Information
  Security Awareness},'' \emph{MIS quarterly}, vol.~34, no.~3, pp. 523--548,
  2010.

\bibitem{moody2018toward}
G.~D. Moody, M.~Siponen, and S.~Pahnila, ``{Toward a Unified Model of
  Information Security Policy Compliance},'' \emph{MIS quarterly}, vol.~42,
  no.~1, pp. 1--50, 2018.

\bibitem{siponen2010neutralization}
M.~Siponen and A.~Vance, ``{Neutralization: New Insights into the Problem of
  Employee Information Systems Security Policy Violations},'' \emph{MIS
  quarterly}, vol.~34, no.~3, pp. 487--502, 2010.

\bibitem{d2014understanding}
J.~D'Arcy, T.~Herath, and M.~K. Shoss, ``{Understanding Employee Responses to
  Stressful Information Security Requirements: A Coping Perspective},''
  \emph{Journal of management information systems}, vol.~31, no.~2, pp.
  285--318, 2014.

\bibitem{2014_Benenson_Defining_Security_Awareness}
N.~Haensch and Z.~Benenson, ``Specifying {IT} security awareness,'' in
  \emph{25th International Workshop on Database and Expert Systems
  Applications, Munich, Germany}.\hskip 1em plus 0.5em minus 0.4em\relax
  {Munich, Germany}: {IEEE}, Sep 2014, pp. 326--330.

\bibitem{WhiteSource2019}
WhiteSource, ``{What are the Most Secure Programming Languages?}'' Mar. 2019,
  {https://tinyurl.com/y2rmfhn7}.

\bibitem{2018_Graziotin_Happy_Developers}
D.~Graziotin, F.~Fagerholm, X.~Wang, and P.~Abrahamsson, ``{What happens when
  software developers are (un)happy},'' \emph{Journal of Systems and Software},
  vol. 140, pp. 32--47, 2017.

\bibitem{WWW_CWE_2019}
\BIBentryALTinterwordspacing
MITRE-Corporation, ``Common~weaknesses~enumeration,'' 2019. [Online].
  Available: \url{https://cwe.mitre.org/}
\BIBentrySTDinterwordspacing

\bibitem{LimeSurvey}
\BIBentryALTinterwordspacing
C.~Schmitz, ``{LimeSurvey v3.17.0},'' Apr. 2020. [Online]. Available:
  \url{\url{https://www.limesurvey.org}}
\BIBentrySTDinterwordspacing

\bibitem{gasiba_zenodo_entire_survey}
\BIBentryALTinterwordspacing
\emph{{Raw Results for the Preliminary Survey on Awareness of Secure Coding
  Guidelines in the Industry}}.\hskip 1em plus 0.5em minus 0.4em\relax Zenodo,
  Oct. 2020. [Online]. Available: \url{https://zenodo.org/record/4075282}
\BIBentrySTDinterwordspacing

\bibitem{thaler2009nudge}
R.~Thaler and C.~Sunstein, \emph{{Nudge: Improving Decisions About Health,
  Wealth, and Happiness}}.\hskip 1em plus 0.5em minus 0.4em\relax {Yale
  University Press}, 2008.

\bibitem{lechner2019security}
U.~Lechner, ``{IT-Security in Critical Infrastructures Experiences, Results and
  Research Directions},'' in \emph{International Conference on Distributed
  Computing and Internet Technology}.\hskip 1em plus 0.5em minus 0.4em\relax
  Springer, 2019, pp. 42--59.

\bibitem{palomba2018diffuseness}
F.~Palomba, G.~Bavota, M.~Di~Penta, F.~Fasano, R.~Oliveto, and A.~De~Lucia,
  ``{On the Diffuseness and the Impact on Maintainability of Code Smells: A
  Large Scale Empirical Investigation},'' \emph{Empirical Software
  Engineering}, vol.~23, no.~3, pp. 1188--1221, 2018.

\bibitem{elkhail2019relating}
A.~A. Elkhail and T.~Cerny, ``{On Relating Code Smells to Security
  Vulnerabilities},'' in \emph{2019 IEEE 5th Intl Conference on Big Data
  Security on Cloud (BigDataSecurity), IEEE Intl Conference on High Performance
  and Smart Computing,(HPSC) and IEEE Intl Conference on Intelligent Data and
  Security (IDS)}.\hskip 1em plus 0.5em minus 0.4em\relax IEEE, 2019, pp.
  7--12.

\bibitem{ENISA_Glossary}
\BIBentryALTinterwordspacing
{European Union Agency for Cybersecurity (ENISA)}, ``{Risk Management
  Glossary},'' Oct. 2020. [Online]. Available:
  \url{https://tinyurl.com/y329vqmb}
\BIBentrySTDinterwordspacing

\bibitem{2018_SAFEcode}
\BIBentryALTinterwordspacing
{Software Assurance Forum for Excellence in Code}, ``{SAFECode - Fundamental
  Practices for Secure Software Development - Essential Elements of a Secure
  Development Life-cycle Program, 3rd Ed.}'' 03 2018. [Online]. Available:
  \url{https://tinyurl.com/y44etrs7}
\BIBentrySTDinterwordspacing

\bibitem{aloraini2019empirical}
B.~Aloraini, M.~Nagappan, D.~M. German, S.~Hayashi, and Y.~Higo, ``{An
  Empirical Study of Security Warnings From Static Application Security Testing
  Tools},'' \emph{Journal of Systems and Software}, 2019.

\bibitem{li2020vulnerabilities}
J.~Li, ``{Vulnerabilities Mapping based on OWASP-SANS: A Survey for Static
  Application Security Testing (SAST)},'' \emph{Annals of Emerging Technologies
  in Computing}, 2020.

\end{thebibliography}
\end{document}